\documentclass[%
 twocolumns,
 notitlepage,
superscriptaddress,
 amsmath,amssymb,
 aps,
 reprint,
pra,
nofootinbib
]{revtex4-1}
\usepackage[dvipsnames]{xcolor}

\usepackage{graphicx}
\usepackage{dcolumn}
\usepackage{bm}
\usepackage{hyperref}
\hypersetup{colorlinks = true, linkcolor = [rgb]{0.19411,0.51882,0.667058}, urlcolor = [rgb]{0.125490,0.29542,0.1647058}, citecolor = [rgb]{0.75882,0.37411,0.14117}}

\usepackage{amsmath}

\usepackage{mathtools}
\usepackage{setspace}

\usepackage{amssymb}

\usepackage{braket}

\graphicspath{{../images/}}
\usepackage[toc,page]{appendix}

\newcommand{\nn}{\nonumber \\}

\usepackage{braket}
\DeclareMathOperator{\tr}{Tr}

\def\>{\rangle}
\def\<{\langle}

\usepackage[bottom]{footmisc}

\usepackage[normalem]{ulem}
\usepackage{makecell}

\begin{document}

\title{Ultimate limits of exoplanet spectroscopy: a quantum approach}

\author{Zixin Huang}
\email{zixin.huang@mq.edu.au}
\affiliation{Centre for Engineered Quantum Systems, School of Mathematical and Physical Sciences, Macquarie University, NSW 2109, Australia}

\author{Christian Schwab}
\affiliation{Centre for Engineered Quantum Systems, School of Mathematical and Physical Sciences, Macquarie University, NSW 2109, Australia}

\author{Cosmo Lupo}
\email{cosmo.lupo@poliba.it}
\affiliation{Dipartimento Interateneo di Fisica, Politecnico di Bari, 70126 Bari, Italy}
\affiliation{INFN, Sezione di Bari, 70126 Bari, Italy}

\begin{abstract}

One of the big challenges in exoplanet science is to determine the atmospheric makeup of extrasolar planets, and to find biosignatures that hint at the existence of biochemical processes on another world. The biomarkers we are trying to detect are gases in the exoplanet atmosphere like oxygen or methane, which have deep absorption features in the visible and near-infrared spectrum. Here we establish the ultimate quantum limit for determining the presence or absence of a spectral absorption line, for a dim source in the presence of a much brighter stellar source. We characterise the associated error exponent in both the frameworks of symmetric and asymmetric hypothesis testing. We found that a structured measurement based on spatial demultiplexing allows us to decouple the light coming from the planet and achieve the ultimate quantum limits.
If the planet has intensity $\epsilon \ll 1$ relative to the star, we show that this approach significantly outperforms direct spectroscopy yielding an improvement of the error exponent by a factor $1/\epsilon$. We find the optimal measurement, which is a combination of interferometric techniques and spectrum analysis.
\end{abstract}

\date{\today}

\maketitle

\section{Introduction}

Humans have long wondered whether other life exists elsewhere in the universe. Signs of life on a distant planet can be detected through the presence of molecules in an exoplanet atmosphere that acts as biomarkers \cite{lederberg1965signs,lovelock1965physical}. If we assume that life uses chemistry for metabolic processes, some of the products, e.g.~oxygen and organic compounds, will accumulate in the atmosphere and lead to non-equilibrium concentrations of these gases. With several thousand exoplanets detected to date, characterising their atmospheres through precision spectroscopy is crucial for our understanding of these worlds. 
To estimate our ability to find biosignatures, this detection task can be best addressed within the theoretical framework of statistical hypothesis testing \cite{lehmann2005testing}. 
The most basic setting involves a binary decision in which the goal is to distinguish between two hypotheses, $H_0$ (the null) and $H_1$ (the alternative). 
There are two ways one can make an error. 
``False positive" (also called type-I error), occurring with probability $\alpha$, is when we accept $H_1$ when the null hypothesis holds. 
Vice versa, ``false negative" (also called type-I error), occurring with probability $\beta$, is the error of accepting the null hypothesis when $H_1$ is true.

For a large enough sample, we can study how $\alpha$ and $\beta$ behave as a function of the sample size $n$. 
In the \textit{symmetric} setting, we are interested in the average probability of error
$ \pi_0 \alpha + \pi_1 \beta$ (where $\pi_0$, $\pi_1$ are the prior probabilities), which is asymptotically bounded by the (quantum) Chernoff bound \cite{chernoff1952measure,PhysRevLett.98.160501} and decays exponentially with increasing $n$.
On the other hand, the costs associated with the two types of errors can be vastly different. As potentially habitable exoplanets are rare, the experimenter may aim at minimising the probability of a false negative (missing the biogas signature), yet, they may be willing to accept false positives as long as the probabilities are below a certain threshold. In this \textit{asymmetric} hypothesis testing setting, the (quantum) Stein lemma \cite{hiai1991proper,ogawa2005strong} provides a bound on
$\beta$ as a function of $n$ and $\alpha$.

The presence of biosignature gases in the exoplanet atmosphere can be inferred from absorption lines in the planet's spectrum \cite{lederberg1965signs,lovelock1965physical}. 
Two main complementary techniques are used to perform atmospheric observation \cite{crossfield2015observations,madhusudhan2019exoplanetary}: transit spectroscopy and direct imaging. 
The majority of atmospheric observations have been accomplished by the transit method: when an exoplanet passes in front of its star in the course of its orbit,  some of the starlight passes through the planet’s atmosphere, and the observed spectrum contains absorption features imprinted on the stellar spectrum.
When the planet is not eclipsing part of the stellar surface from our vantage point, we only observe the star's spectrum. Measuring the difference of the stellar spectrum with and without the planet’s signature imprinted on it enables us to characterise the absorption spectrum of the planet’s atmosphere \cite{burrows2014spectra}. 
However, since the absorption of the planet’s atmosphere only affects a tiny portion of the starlight, determined by the relative area of the projection of the planet’s atmosphere compared to the stellar surface, detecting such absorption features presents a severe challenge to transit spectroscopy \cite{crossfield2015observations}.

Spectroscopy by direct imaging -- where the star-planet system is (at least partially) spatially resolved -- is a powerful complement to the transit method. However, due to diffraction effects of the telescope optics, and turbulence in the Earth’s atmosphere for ground-based observations, the image of a point-like object (star) is not a point, but will have a finite size characterised by the point-spread function (PSF).
Because of the small angular separation between planets and their host stars, their PSFs may overlap in the focal plane of the imaging system. In particular, the weak light received from the planet is typically buried in the wings of the star’s PSF.
The use of coronagraphs allows astronomers to block out the light of the star, and create an area around the centre of the star’s image where the exoplanet can be detected above the background produced by the starlight. However, because of the extreme contrast ratio between planet and star (ranging from $10^{-3}$ to $10^{-10}$ for hot Jupiter to Earth-like analogues \cite{birkby2018spectroscopic}), this presents the greatest challenge in direct imaging.
It is noteworthy that, for exoplanets close enough to us, the planet itself is not necessarily dim compared to other astronomical objects, and if the stellar and planetary light could be separated perfectly, spectroscopy of the planet could be readily performed~\cite{fischer2014protostars}. 

In this paper, we describe how to use techniques from quantum imaging to boost the sensitivity of exoplanet atmospheric spectroscopy. We use tools from quantum information theory to quantify the ultimate limits, as expressed by the quantum Chernoff bound and the quantum Stein lemma. 
Then, we show how the star and planet's full spatiospectral information can be 
extracted using a relatively simple, linear optical measurement, consisting of SPAtial DE-multiplexing (SPADE) followed by spectroscopy. 
It has been shown that SPADE can enhance the 
estimation of the separation between two point-like incoherent sources~\cite{PhysRevX.6.031033,PhysRevLett.117.190801,PhysRevLett.117.190802,tsang2019resolving} and the discrimination between sources with different shapes and features~\cite{Tsang_NJP,Modern,PhysRevLett.124.080503,Lvovsky,grace2021identifying,HKrovi}, including in particular the problem of detecting exoplanets~\cite{PhysRevLett.127.130502}.
Here we use SPADE to first decouple the light coming from the planet from that of the star, and then proceed with a spectral analysis. We show that this approach is optimal in the sense that it saturates the ultimate quantum limits to spectroscopy.

\begin{figure}[t]
\includegraphics[trim = 0cm 0.0cm 0cm 0cm, clip, width=1.0\linewidth]{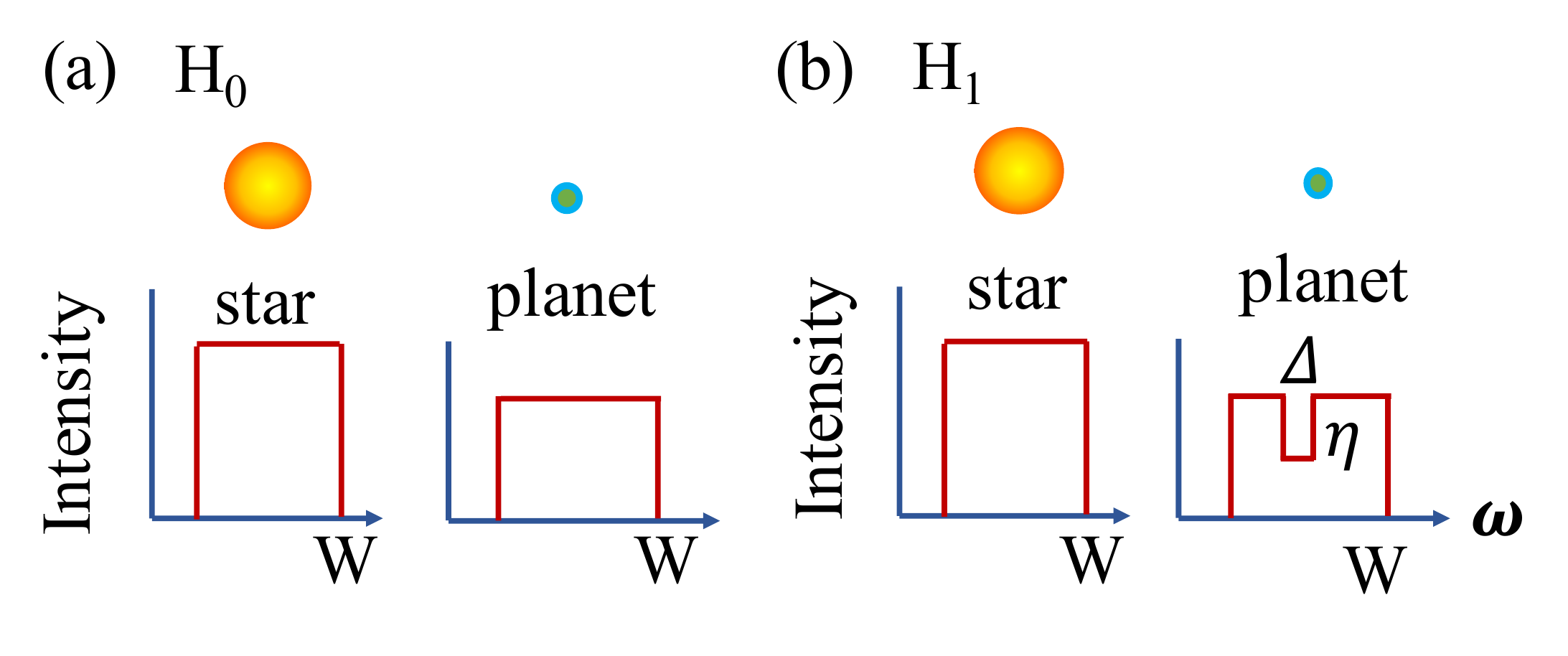} 
\caption{(a) The spectrum of the star and planet under $H_0$: no absorption line is present. (b) Under $H_1$, 
an absorption line is present in the exoplanet's atmosphere.
}
\label{f:schematic1}
\end{figure}

\begin{figure}[t]
\includegraphics[trim = 0cm 1.0cm 0cm 0cm, clip, width=0.9\linewidth]{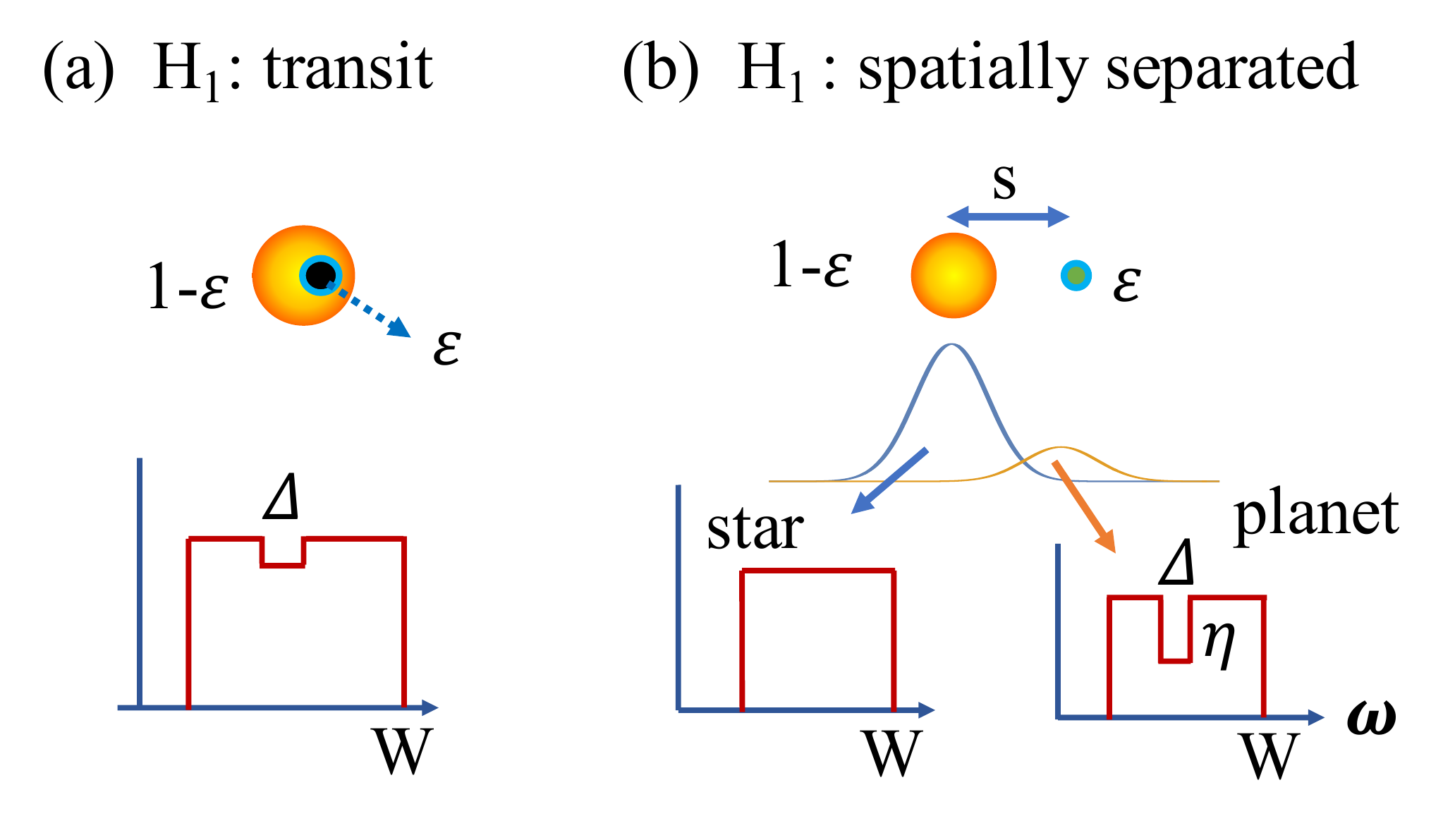} 
\caption{\label{f:schematic2} Two most common physical implementations of exoplanet spectroscopy: (a) the transit method spectroscopy under $H_1$: exoplanet passes in front of their star, and an absorption line is imprinted onto the exoplanet's atmosphere. The relative intensity of the planet to the star is $\epsilon/(1-\epsilon) \ll1$, and the absorption coefficient of the planet's atmosphere of the wavelengths of interest is $\eta$. (c) The star-planet system is spatially separated, and the relative intensity is $\epsilon/(1-\epsilon) \ll1$. The absorption coefficient of the planet's atmosphere is $\eta$. In general the values of $\epsilon$ and $\eta$ can be different for the transit and spatially separated cases~\cite{deming2018characterize}, because the stellar light goes through different thicknesses of the exoplanet's atmosphere. }
\end{figure}

\section{Summary of the results and structure of the paper}

In this work, we quantify the ultimate quantum limit for detecting the presence of a spectral absorption line. The metrics of interest are the error exponents of the (quantum) Chernoff bound and the (quantum) Stein lemma.
These quantities are summarised in Table~\ref{tab1}, where we show the scaling in terms of the relative intensity $\epsilon$ of the planet ($\epsilon \ll 1$), and in the parameter $|\kappa|^2$ ($|\kappa|^2 \lesssim 1$) which quantifies the overlap of the PSFs.
The main results of our work are: 1) we show that the transit method and direct imaging have error exponents that scale as $\epsilon^2$, whereas the optimal quantum method scales as $\epsilon$; 2) we find the linear-optical measurement that saturates these error bounds, and 3) given a star-planet system, we allow the experimenter to calculate the achievable error bounds as a function of the number of photons received, taking into account finite-size effects.

\begin{table*}[t]
\begin{center}
\renewcommand{\arraystretch}{1.7}
\setlength{\tabcolsep}{18pt}
\begin{tabular}{ ccccc }
\hline
               & Transit method & Direct imaging & Quantum limit & SPADE \\ 
\hline
\makecell{Error exponent} & $O(\epsilon^2 )$ & $ |\kappa|^{-8} O(\epsilon^2 )$ & $  \left( 1-|\kappa|^2 \right) O( \epsilon )$ & $\left( 1-|\kappa|^2 \right) O( \epsilon ) $ \\
 \hline
\end{tabular}
\end{center}
\caption{
\label{tab1}Summary of the results. The parameters of interest are: $\epsilon$ is the relative intensity of the planet
and $|\kappa|^2$ is the overlap of the PSFs of the star and planet. We use Gaussian PSF as an example, and in this case, $\kappa$ is real-valued. In general we expect the same scaling with $\epsilon$ and $|\kappa|^2$ for other PSF model.}
\end{table*}

\color{black}

The structure of the paper is as follows. In Sec.~\ref{sec:theory} we introduce the theoretical framework for determining the error probabilities in the hypothesis test.
In Sec.~\ref{Sec:diff} we introduce the model of diffraction-limited telescope.
In Sec.~\ref{sec:results}, we present our results on spectroscopy via (a) the transit method, (b) via direct imaging;
we compare these to the ultimate quantum limit and show that a quadratic improvement is obtained by using the optimal measurement. 
The hypothesis testing is affected by having a finite-size $n$. 
For the asymmetric test we characterise how the false-negative probability $\beta$ behaves in this case in Sec.~\ref{sec:finite}; this will be particularly applicable to the scenarios where we are limited by the number of photons received.

\section{Theoretical toolbox}
\label{sec:theory}

Taking into account the quantum nature of light, the two hypotheses are associated with a pair of quantum states $\rho_0$ (for the null hypothesis) and $\rho_1$ (for the alternative hypothesis). 
Such states are used to represent the signal coming from the star-planet system.
In our analysis, we consider the regime of highly attenuated signals, where at most one photon is detected
within a detection window.
The experimenter collects and measures a number $n$ of identical signals from the star-planet system,
which in turn are described by the tensor product state $\rho_j^{\otimes n}$. The goal is to determine whether $j=0$ or $j=1$.
Below we formally describe this problem in two alternative ways, namely symmetric and asymmetric hypothesis testing, and derive bounds on the probability of error.

\subsection{Symmetric hypothesis testing}

In the symmetric setting, given the priors $\pi_0,\pi_1$, with $\pi_0 +\pi_1 =1$, the goal is to find the measurement strategy that minimises the average error probability
$P_{e,n} = \pi_0 \alpha + \pi_1 \beta$.
The Helstrom bound is a lower bound on the error probability~\cite{Helstrom_book,Holevo1978}, 
\begin{align}
P_{e,n} 
\geq \frac{1}{2}\left( 1- \| \pi_1 \rho_1^{\otimes n} - \pi_0 \rho_0 ^{\otimes n} \| \right) \, .
\end{align}
The bound is tight, and can be saturated for any value of $n$ with the so-called \textit{Helstrom measurement}, which is the projection into the positive and negative parts of the difference $\pi_1 \rho_1^{\otimes n} - \pi_0 \rho_0 ^{\otimes n}$.
The Helstrom bound may be difficult to compute for arbitrary values of $n$. However, for large $n$, we expect that the probability of error decreases exponentially with $n$, with error exponent 
\begin{align}
\xi = - \lim_{n\to\infty} \frac{1}{n}\log 
P_{e,n}
\, .
\end{align}
The quantum Chernoff bound gives an upper bound on the asymptotic error exponent~\cite{qchernoff,PhysRevLett.98.160501}
\begin{align}
\xi & \leq \xi_\text{QCB} := - \log 
\min_{0 \leq t \leq 1}  \text{Tr} \left( \rho_0^t \rho_1^{1-t} \right)
\end{align}
Note that this bound is universal, in the sense that it depends only on the states $\rho_0$ and $\rho_1$. It already implicitly accounts for the optimisation over all possible measurements allowed by the principles of quantum mechanics. 
However, the quantum Chernoff bound does not give the explicit form of the optimal measurement.
In general, the optimal measurement may be highly non-local.

Consider instead a local measurement $\mathcal{M}$, applied to each instance of the unknown state $\rho_j$, which yields as an outcome a random variable $x$ with associated probability distribution $p_j^{\mathcal{M}}(x)$.
Once a measurement is performed, the statistical test is entirely classical. In this case, the classical Chernoff bound gives us an upper bound on the error exponent associated to this particular measurement~\cite{chernoff1952measure}.
For example, if the outcome $x$ is a continuous variable we have
\begin{align}\label{eq:ccb}
\xi^\mathcal{M}_\text{CCB}(\mathcal{M}) = - \log{ 
\min_{0 \leq t \leq 1} \int dx 
\left[ p_0^{\mathcal{M}}(x)\right]^t 
\left[ p_1^{\mathcal{M}}(x) \right]^{1-t} 
} 
.
\end{align}
For a generic measurement $\xi_\text{CCB}(\mathcal{M}) \leq \xi_\text{QCE}$, with equality holding only if the measurement is optimal.

\subsection{Asymmetric hypothesis testing}\label{ssec:asym}

In the asymmetric setting \cite{Petz1991,Ogawa2000, PhysRevLett.119.120501} we require the type-I error to be less than $1/2$, and look for the optimal measurement that minimises the type-II error.
Also in this case, we expect that $\beta$ decays exponentially with increasing number of copies $n$, with asymptotic error exponent
\begin{align}
\vartheta = - \lim_{n\to\infty} \frac{1}{n}\log 
\beta 
\, .
\end{align}
The quantum Stein lemma allows us to compute a universal upper bound on this exponent 
~\cite{Petz1991,Ogawa2000}, 
\begin{align}
\vartheta \leq \vartheta_\text{QRE} :=  D(\rho_0 \| \rho_1) \, ,
\end{align}
where 
\begin{align}
D(\rho_0 \| \rho_1) = \text{Tr}[\rho_0 (\log\rho_0 - \log \rho_1)]
\end{align}
is the quantum relative entropy~\cite{wilde2013quantum}.
Note that the relative entropy is the expectation value of the logarithmic likelihood ratio $\log\rho_0 - \log \rho_1$.
As for the quantum Chernoff bound this holds for all measurements. It is also tight, as there exists an optimal measurement that asymptotically saturates it.

Consider now a local measurement $\mathcal{M}$, which outputs $x$ with probability $p_j^{\mathcal{M}}(x)$ when applied to the state $\rho_j$. The probability of error still decays exponentially with $n$, but with an asymptotic exponent that is bounded by the classical relative entropy is~\cite{Cover2006_book}
\begin{align}
\vartheta^\mathcal{M}_\text{CRE} := D_c (p^\mathcal{M}_0 \| p^\mathcal{M}_1) = \int dx \, p^\mathcal{M}_0(x) \log \frac{ p^\mathcal{M}_0(x) }{ p^\mathcal{M}_1(x) } \, .
\end{align}
In general, $D_c (p^\mathcal{M}_0 \| p^\mathcal{M}_1) \leq D(\rho_0 \| \rho_1)$ {for any $\mathcal{M}$}. If equality holds, the measurement is asymptotically optimal, as
the error exponent matches the quantum relative entropy.

\section{Diffraction-limited telescope}\label{Sec:diff}

Consider a diffraction-limited telescope used to create an image of the star-planet system in the far field and in the paraxial regime. The telescope is characterised by its PSF. The star and the planet lay in the object plane, and the telescope creates a focused image on the image plane~\cite{goodman2008introduction}.
To provide a quantum description of the optical field, we ignore polarisation and work with a scalar field defined on the image plane, where light is collected and measured. We denote $a_{x,\omega}$, $a_{x,\omega}^\dag$ as the Bosonic canonical operators that annihilate and create a photon at point $x$ on the image plane with frequency $\omega$.
The image of a quasi-monochromatic, point-like object, located at point $y$ on the object plane, is associated with the following canonical operators on the image plane:
\begin{align}
a_{y,\omega} & = \int dx \, \psi(x - y/M)^* \, a_{x,\omega} \, , \\
a_{y,\omega}^\dag & = \int dx \, \psi(x - y/M) \, a_{x,\omega}^\dag \, ,
\end{align}
where $\psi$ is the PSF, ${}^*$ denotes complex conjugation, and $M$ is the magnification factor.
To simplify the notation, in the following we set $M=1$ without loss of generality \cite{tsang2019resolving}.

This work focuses on the regime of highly attenuated signals, where the probability of receiving more than one photon from the star is negligible. Furthermore, both the star and the planet are assumed to be point-like emitters.
In this scenario, if the telescope is centred on the star, a single, monochromatic photon arriving on the image plane is described by the state 
\begin{align}
    |\psi_{0,\omega}\rangle = \int dx \, \psi(x) \, a_{x,\omega}^\dag |0\rangle \, ,
\end{align}
where $|0\rangle$ is the vacuum.
The planet, which on the object plane is separated from the star by a transverse distance $s$, yields on the image plane the single-photon state
\begin{align}
    |\psi_{s,\omega}\rangle = \int dx \, \psi(x-s) \, a_{x,\omega}^\dag |0\rangle \, .
\end{align}

Note that these two states have a non-zero overlap, 
\begin{align}\label{kappadef}
    \kappa = \langle \psi_{0,\omega} | \psi_{s,\omega} \rangle 
    = \int dx \, \psi(x)^* \psi(x-s) \neq 0 \, .
\end{align}
This is due to diffraction through the finite aperture of the telescope.
For example, for a Gaussian PSF,
\begin{align}
    \psi(x) = \frac{1}{(2\pi \sigma^2)^{1/4}} \,  e^{-x^2/4\sigma^2}
    \, ,
\end{align}
the parameter $\kappa$ is real with
$|\kappa|^2 = e^{-s^2/4\sigma^2}$.
Note that the quantity $\sigma$ plays the role of the Rayleigh length~\cite{goodman2008introduction}. 
The two states on the image plane become well separated when $s \gg \sigma$.
Otherwise, in the sub-Rayleigh regime, when $s \lesssim \sigma$, the two states have substantial overlap and are hard to discriminate.
This in turn limits the effectiveness of direct imaging methods for spectroscopy.

For non-monochromatic incoherent sources, we need to sum over contributions from different frequencies. Therefore, including the spectral degrees of freedom, the states of the photon coming from the planet and star are
\begin{align}
    \rho_\text{star} & = \int d\omega \, S_\text{star}(\omega) \, | \psi_{0,\omega} \rangle  \langle \psi_{0,\omega} | \, , \\
    \rho_\text{planet} & = \int d\omega \, S_\text{planet}(\omega) \, | \psi_{s,\omega}  \rangle \langle \psi_{s,\omega} | \, ,
\end{align}
where $S_\text{star}$, $S_\text{planet}$ are the spectral densities of the two sources.
To facilitate computations on these states, it is convenient to separate the spectral and spatial degrees of freedom by introducing a tensor product structure. We define 
\begin{align}
    |x\omega \rangle = |x \rangle \otimes |\omega\rangle 
    := a_{x,\omega}^\dag |0\rangle \, ,
\end{align}
which is the state of a photon associated with a particular position $x$ on the image plane and a well-defined frequency $\omega$.
With this notation, the above states read
\begin{align}
    \rho_\text{star} & = \int d\omega \, S_\text{star}(\omega) \, |\omega\rangle \langle \omega| \otimes  \, | \psi_0 \rangle  \langle \psi_0 | \, , \\
    \rho_\text{planet} & = \int d\omega \, S_\text{planet}(\omega) \, |\omega\rangle \langle \omega| \otimes | \psi_s \rangle \langle \psi_s | \, ,
\end{align}
where
\begin{align}
    | \psi_0 \rangle  = \int dx \, \psi(x) \, |x\rangle \, , \quad
    | \psi_s \rangle  = \int dx \, \psi(x-s) \, |x\rangle \, .
\end{align}

\section{The Model}

In the previous section, we have introduced the main theoretical tools that we will use to estimate the error probability and compare different measurement strategies. 
Now we are ready to introduce an explicit model for the states $\rho_0$ and $\rho_1$ corresponding to the null and alternative hypotheses.
We assume that the presence of the exoplanet is known, and the goal is to detect a molecular absorption line in the planet's atmosphere. 
In a typical stellar spectroscopy experiment, the experimenter would measure the spectrum of the star alone, and contrast it against the spectrum of the star-planet system.  Since the stellar source is typically bright and stable \cite{deming2018characterize}, the stellar spectrum can be well-characterised.

Since we only wish to examine the difference between the star and the planet's spectra, for simplicity, under $H_0$, we can model a photon coming from the star as having a flat spectrum: in any case,  we can account for any variation via renormalisation \cite{massey2013astronomical,eaton2007tennessee,pirzkal2004grapes}.  Under the same hypothesis, there is no absorption line, and a photon coming from the planet's atmosphere has the same spectrum.
The state of a photon coming from the star is written as
\begin{align}\label{rhostar}
\rho^\text{star} & = \frac{1}{W} \int_0^{W} d\omega \ket{\omega}\bra{\omega} \otimes \ket{\psi_0}\bra{\psi_0} \\
& =: \gamma_0 \otimes \ket{\psi_0}\bra{\psi_0} \, ,
\end{align}
where $\gamma_0$ only depends on the frequency degree of freedom.
Under $H_0$, the photon coming from the planet has the same spectrum, therefore it is represented by the state
\begin{align}\label{rhoplanet0}
\rho^\text{planet}_0 &= \frac{1}{W} \int_0^{W} d\omega \ket{\omega}\bra{\omega} \otimes \ket{\psi_s}\bra{\psi_s} \\
               & =: \gamma_0 \otimes \ket{\psi_s}\bra{\psi_s} \, .
\end{align}
The spectral distributions associated with hypothesis $H_0$ are schematically shown in Fig.~\ref{f:schematic1}(a).

Under $H_1$, a photon coming from the planet shows an absorption line centred at $\omega_0$ of width $\Delta < W$. Realistically, the absorption profile is Gaussian due to the Doppler shift of the gaseous molecules, but the only parameter of importance is the total area under integration. Therefore, for mathematical simplicity we will model the profile to be rectangular; the width of the band is known as the ``equivalent width" \cite{sousa2007new,sousa2008spectroscopic}.
Thus, the density matrix of a photon coming from the planet is
\begin{align}\label{rhoplanet1}
\rho^\text{planet}_1 & = \frac{1}{W - \eta \Delta }\bigg[ \int_0^{\omega_0 - \Delta /2} d\omega  
         + (1-\eta) \int_{\omega_0-\Delta /2}^{\omega_0 + \Delta/2} d\omega \nn  
         & \phantom{=}~ +  \int_{\omega_0 + \Delta/2}^W  d\omega  \bigg] \ket{\omega}\bra{\omega} 
         \otimes \ket{\psi_s}\bra{\psi_s} \\
         & =: \gamma_1 \otimes \ket{\psi_s}\bra{\psi_s}\, ,
\end{align}
where $\eta$ is the absorption coefficient and the operator $\gamma_1$ depends only on frequency.
The spectral distributions for $H_1$ are shown in Fig.~\ref{f:schematic1}(b).

We are now in the position of writing the quantum states associated with the two hypotheses. For the null hypothesis ($H_0$), that there is no absorption line, we have
\begin{align}
    \rho_0 = (1-\epsilon) \rho^\text{star} + \epsilon \rho^\text{planet}_0 \, ,
\end{align}
whereas for the alternative hypothesis ($H_1$), that there is an absorption line, \begin{align}
    \rho_1 = (1-\epsilon) \rho^\text{star} + \epsilon \rho^\text{planet}_1 \, ,
\end{align}
with $\rho^\text{star}$, $\rho^\text{planet}_0$ and $\rho^\text{planet}_1$ as in Eqs.~(\ref{rhostar})-(\ref{rhoplanet1}), and $\epsilon$ is the relative intensity of the light coming from the planet.

\section{Results} \label{sec:results}

\subsection{Transit spectroscopy}

First, we will examine transit spectroscopy, which is the most widely used method in astronomy \cite{deming2018characterize}.
The method refers to the setting where the planet passes in front of the star, as schematically depicted in Fig.~\ref{f:schematic2}(a). 
In our model, this corresponds to the case where $s=0$ (i.e.~$\kappa = 1$), which implies
\begin{align}
    \rho_0 & = \gamma_0 \otimes \ket{\psi_0}\bra{\psi_0} \, , \qquad
 \rho_1 = \gamma_1 \otimes \ket{\psi_0}\bra{\psi_0} \, .
\end{align}
Therefore, the spatial degrees of freedom are factored out, and information is only carried by frequency. 
Note that the state $\gamma_0$ and $\gamma_1$ commute as they are both diagonal in the frequency basis. This implies that the quantum Chernoff bound coincides with the classical Chernoff bound obtained by a projective measurement in the frequency basis. For the same reason, the quantum relative entropy equals the classical relative entropy corresponding to such a measurement.
In other words, spectroscopy by the transit method is already optimal when the PSF of the planet completely overlaps with that of the star.

For a Gaussian PSF, we can compute an explicit expression for the Chernoff exponent. For $\epsilon \ll 1$. We obtain (see Appendix~\ref{sec:transit})
\begin{align}
\xi_\text{QCB} = \xi^\text{transit}_\text{CCB}
& = \frac{ \epsilon^2 \eta^2 (W-\Delta )\Delta }{8 (W-\eta\Delta )^2} 
+ O(\epsilon^3) \, .
\label{TranQCE}
\end{align}
Similarly, for asymmetric hypothesis testing the error exponents read
\begin{align}
\vartheta_\text{QRE} =
\vartheta^\text{transit}_\text{CRE}
& = \frac{ \epsilon^2 \eta^2 (W-\Delta ) \Delta}{2 (W-\eta\Delta )^2} 
+ O(\epsilon^3) \, .
\label{TranQRE}
\end{align}

We note that all the exponents scale
quadratically in $\epsilon$. This formally expresses the challenges of using transit spectroscopy: because the light of the dim planet is mixed in with the star's, the two spectra are difficult to distinguish.
We also note that at the leading order the exponents for symmetric and asymmetric hypothesis tests are identical up to a constant.

\subsection{Classical limits to direct imaging}

In the previous section, we have considered the case of complete overlap between the star and planet's PSFs.
We now consider the case of partial overlap. First, we establish a classical benchmark for direct imaging applied to this setup. 
Partial overlap means~$s>0$ ($\kappa <1$) and is schematically shown in Fig.~\ref{f:schematic2} (b). 

Direct imaging (DI) consists of measuring the intensity of the field as focused on the image plane, i.e., pixel-by-pixel photon detection.
This approach is readily extended to spectroscopy if a spectral analysis is performed on each pixel.
We can calculate the probability distribution of detecting a photon of frequency $\omega$ at position $x$ on the image screen.
For $H_j, j \in \{0,1\}$,
\begin{align}\label{eq:DI_pdf}
p^\text{DI}_j(x\omega) 
& = \langle x\omega | \rho_j | x\omega \rangle \nn
& = (1-\epsilon) \bra{x\omega}\rho^\text{star}\ket{x\omega}  
+ \epsilon \bra{x\omega}\rho^\text{planet}_j\ket{x\omega} \, .
\end{align}

\begin{figure}[t]
\includegraphics[trim = 0cm 0.0cm 0cm 0cm, clip, width=1.0\linewidth]{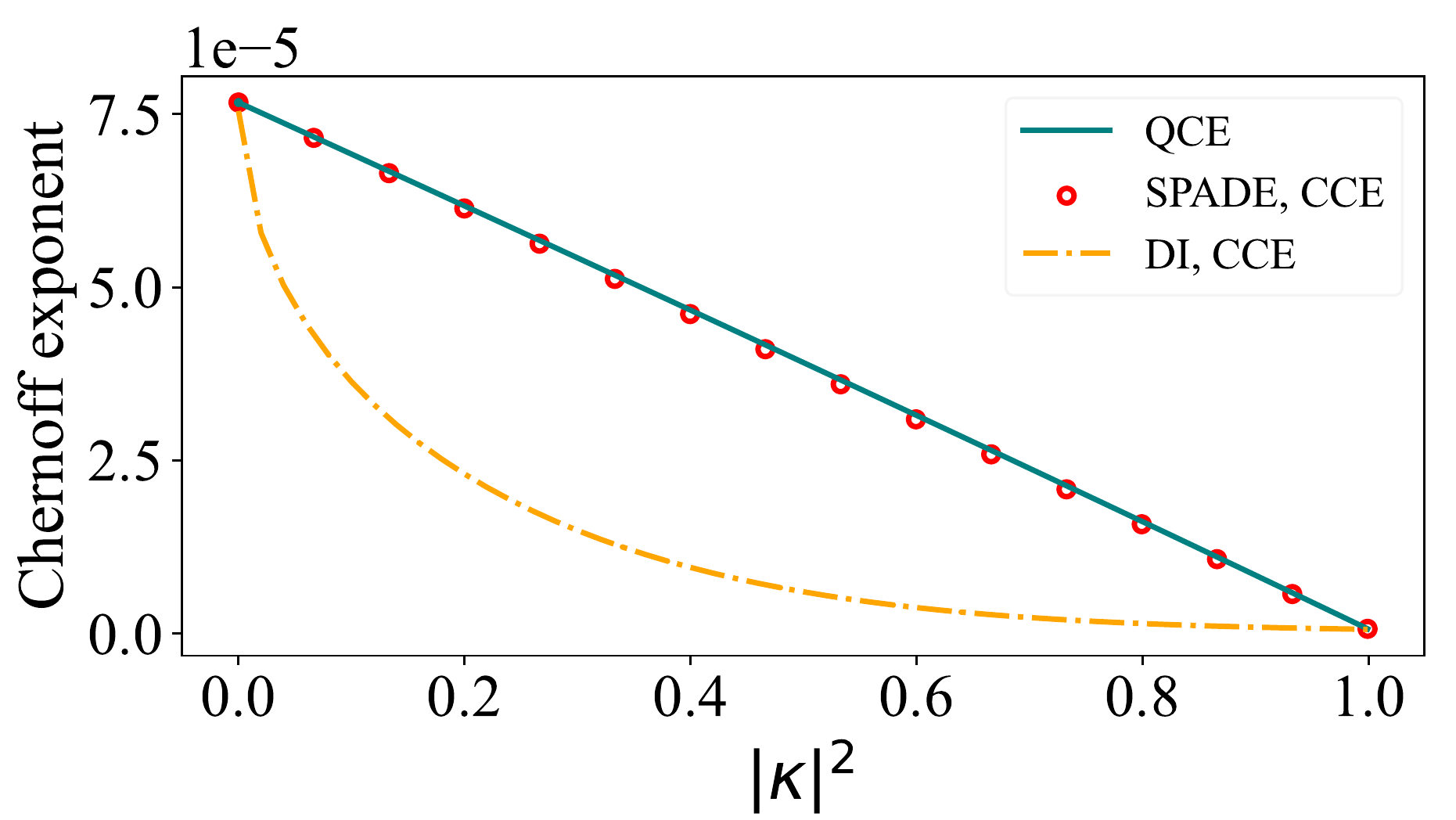} 
\includegraphics[trim = 0cm 0.0cm 0cm 0.3cm, clip, width=1.0\linewidth]{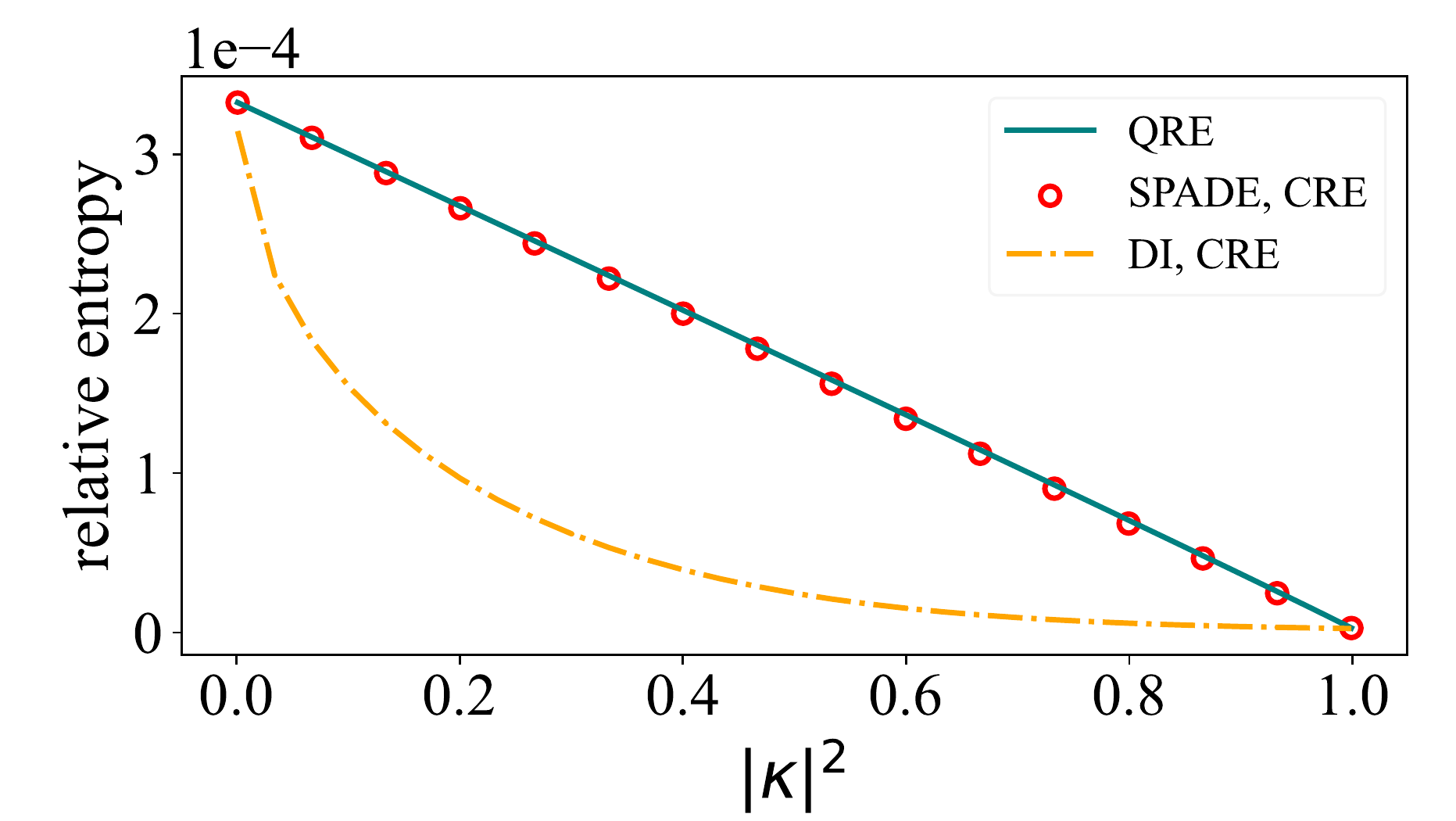} 
\caption{
\textbf{Top figure:} error exponents for symmetric hypothesis testing, plotted vs the overlap $|\kappa|^2$.
Green solid line: the ultimate quantum limit, expressed by the quantum Chernoff exponent (QCE) (calculated numerically). Yellow dotted-dashed line: the classical Chernoff exponent (CCE) for direct imaging (DI) for a Gaussian PSF, calculated numerically from Eq.~(\ref{eq:ccb_di}).
The red circles indicate the classical Chernoff exponent
for a nearly optimal SPADE-assisted spectral measurement in Eq.~\eqref{eq:measurement1}-\eqref{eq:measurement2}. %
\textbf{Bottom figure:} error exponents for asymmetric hypothesis testing.
Green solid line: the ultimate quantum limit, expressed by the quantum relative entropy (QRE).
Yellow dotted-dashed line: the classical relative entropy (CRE) of direct imaging for a Gaussian PSF, calculated numerically from Eq.~\eqref{eq:Dc}.
The red circles show the classical relative entropy for the nearly optimal SPADE-assisted spectral measurement in Eq.~\eqref{eq:measurement1}-\eqref{eq:measurement2}. %
The parameters of the model are $\epsilon = 0.01 $, $W=1$ , $\Delta = 0.2W$, $\eta=0.5$.
}
\label{f:qre_vs_cre_main}
\end{figure}

To estimate the error probabilities we compute the Chernoff error exponent
\begin{align} \label{eq:ccb_di}
\xi^\text{DI}_\text{CCB} & = - \log 
\min_{0\leq t \leq 1} \iint dx \, d\omega 
\left [p^\text{DI}_{0}(x\omega) \right]^t \left[ p^\text{DI}_{1}(x\omega) \right]^{1-t} 
\, , 
\end{align}
and the classical relative entropy
\begin{align}\label{eq:Dc}
\vartheta^\text{DI}_\text{CRE} 
& = \iint dx \, d\omega~ p^\text{DI}_{0}(x\omega) \log \frac{ p^\text{DI}_{0}(x\omega) }{ p^\text{DI}_{1}(x\omega) }  \, .
\end{align}
These error exponents can be computed explicitly, for example for a Gaussian PSF. 
For a dim planet ($\epsilon \ll 1$) very close to the star ($s \lesssim \sigma$).
In this limit, the error exponents read
\begin{align}
\xi^\text{DI}_\text{CCB} & = \frac{ \epsilon^2 \eta^2 |\kappa|^{-8} (W-\Delta ) \Delta}{8 (W-\eta\Delta)^2 } + O(\epsilon^3) \, , \\
\vartheta^\text{DI}_\text{CRE} 
&= \frac{ \epsilon^2 \eta^2 |\kappa|^{-8} (W-\Delta )\Delta  }{2 (W-\eta\Delta)^2 } + O(\epsilon^3) \, .
\label{CRE_DI}
\end{align}
These expressions for the error exponents generalise those of Eq.~(\ref{TranQCE})-(\ref{TranQRE}) to the region where $s > 0$ (i.e.~$\kappa <1$), and show the same quadratic scaling in $\epsilon$. 
As expected, by increasing the separation $s$, the images of the star and the planet become more distinguishable, as well as their spectra. 
Note that the above expansions hold for values of $s/\sigma$ up to $\approx 1.3$, corresponding to $\kappa \approx 0.8$.
The two error exponents are shown in Fig.~\ref{f:qre_vs_cre_main}, where the yellow, dot-dashed lines plot them vs the overlap parameter $\kappa \in [0,1]$.

\subsection{Quantum limits of spectroscopy}

The most interesting setting is when the star and planet create, due to diffraction, partially overlapping images on the image plane. 
As previous work on sub-Rayleigh resolution suggests~\cite{PhysRevX.6.031033,PhysRevLett.127.130502,grace2021identifying}, direct imaging may not be the optimal method to discriminate the absorption line, especially when the planet and the star's PSFs partially overlap.
In this section, we use the notion of quantum Chernoff bound and quantum relative entropy to compute the best possible error exponents allowed by quantum mechanics, corresponding to the application of a globally optimal measurement.

As noted above the spatial wave functions $\ket{\psi_0}$ and $\ket{\psi_s}$ are in general not orthogonal. They have a substantial overlap if their separation $s$ is comparable or smaller than the Rayleigh length.
To calculate the error exponents, it is convenient to write the states $\rho_0$, $\rho_1$ using a basis set that spans the Hilbert space of the spatial degree of freedom. One convenient choice is to orthogonalise the vectors $\ket{\psi_0}$, $\ket{\psi_s}$ into
\begin{align}
\ket{e_1} = \ket{\psi_0} \, , \qquad
\ket{e_2} = \frac{\ket{\psi_s} - \kappa\ket{\psi_0}}{\sqrt{1-|\kappa|^2}} \, ,
\end{align}
where the spatial overlap $\kappa  = \braket{\psi_0|\psi_s}$ is the same as for monochromatic light in Eq.~(\ref{kappadef}).
Combining the spatial and spectral degrees of freedom, the density matrices associated with the two hypotheses are
\begin{align}  
\rho_0 & = \sigma_0 
\otimes 
[ (1-\epsilon (1-|\kappa|^2) ] \ket{e_1}\bra{e_1} + \epsilon (1-|\kappa|^2) \ket{e_2}\bra{e_2} \nn
& \phantom{=}~+\epsilon \kappa\sqrt{1-|\kappa|^2} \, 
(\ket{e_1}\bra{e_2} + \text{h.c.} ) ] \, ,
\label{eq:nondiag_main}
\end{align}
and
\begin{align}  
\rho_1 & = \left[ (1-\epsilon) \sigma_0 + \epsilon |\kappa|^2 \sigma_1 \right] \otimes \ket{e_1}\bra{e_1} \nn
& \phantom{=}~ + \epsilon (1-|\kappa|^2) \sigma_1 \otimes \ket{e_2 }\bra{e_2} \nn
& \phantom{=}~+ \epsilon \kappa \sqrt{1-|\kappa|^2} \, \sigma_1 \otimes 
( \ket{e_1}\bra{e_2} + \text{h.c.} )
\, .
\label{eq:nondiag_main_2}
\end{align}
Starting from this representation, we can compute the quantum relative entropy.
In the limit that $\epsilon \ll 1$, we obtain
\begin{align} \label{QRE-s}
\vartheta_\text{QRE}
 = \frac{ \epsilon (1-|\kappa|^2) \Omega}{W} + O(\epsilon^2) \, , 
 \end{align}
with
\begin{align} \label{Omegadef}
\Omega = W \log \left(\frac{W-\eta\Delta}{ W}\right)- \Delta  \log (1-\eta ) \, ,
\end{align}
and the second-order term is given in Appendix~\ref{AppA2}.
Due to the optimisation in $t$, the quantum Chernoff exponent for this case is difficult to compute analytically but is straightforward to calculate numerically.
Both error exponents are shown in Fig.~\ref{f:qre_vs_cre_main}, see the solid, blue lines. The two quantities share a similar behaviour as a function of $|\kappa|^2$.

The above expression for $\vartheta_\text{QRE}$ is valid for all values of $\kappa$.
Notably, the leading term in Eq.~(\ref{QRE-s}) scales linearly in $\epsilon$.
This term is multiplied by the factor $(1-|\kappa|^2)$, which is non-zero as long as the overlap between the two PSFs is incomplete.
In particular, when $|\kappa|^2 = 1$ (i.e.~$s = 0$) the sources' PSFs completely overlap and this leading term vanishes. In fact, in this limit, we recover the error exponent for the transit method.
On the other hand, we see that direct imaging (yellow dot-dashed lines in Fig.~\ref{f:qre_vs_cre_main}) rapidly becomes sub-optimal as soon as $|\kappa|^2 < 1$ (i.e.~$s > 0$), and clearly shows a different scaling with $\kappa$. Only in the limit of well-separated sources when $\kappa = 0$ (i.e.~$s \gg \sigma$) direct imaging becomes optimal again. 
\color{black}

\begin{figure}[t]
\includegraphics[trim = 0cm 0.0cm 0cm 0cm, clip, width=0.9\linewidth]{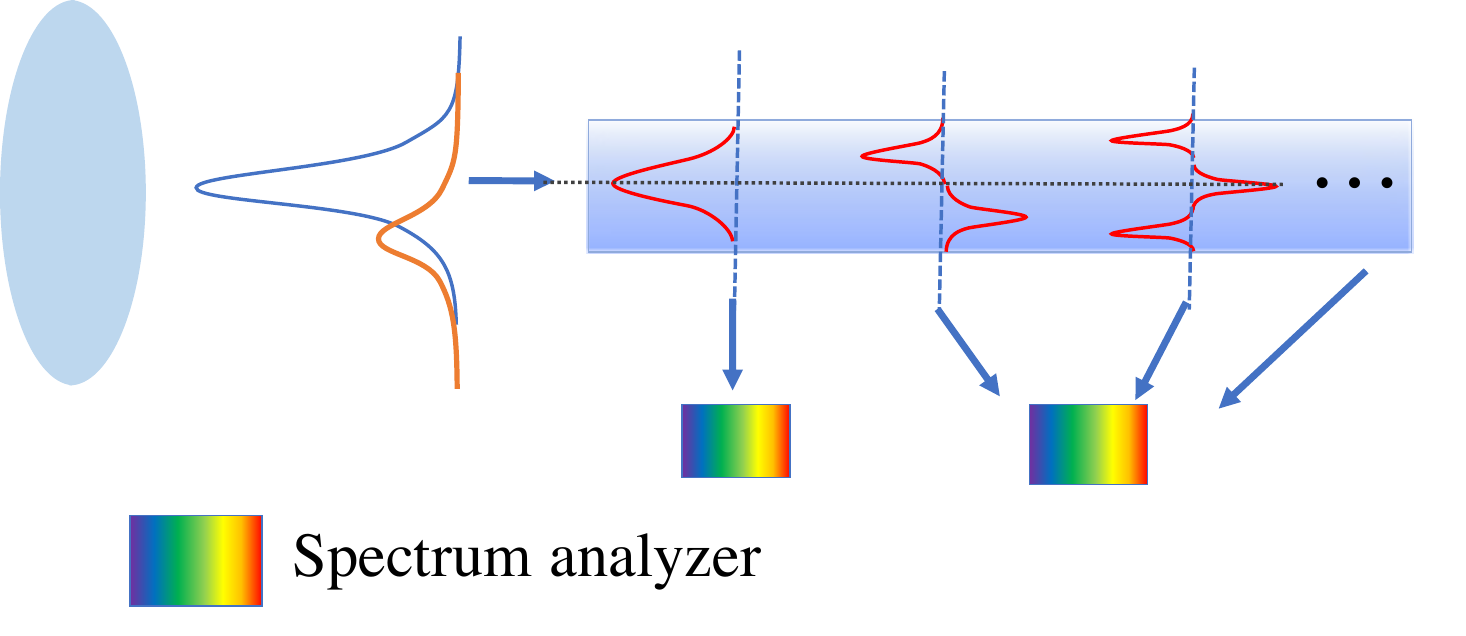} 
\caption{\label{f:measurement}A multimode waveguide can be used as Hermite-Gaussian mode sorter \cite{PhysRevX.6.031033,PhysRevLett.117.190801}. Light entering the device can be described as a combination of Hermite-Gaussian modes, each input mode is demultiplexed and sent to a separate mode, which is coupled into a spectrum analyser.
The star is aligned with the centre of the waveguide. At the output, the light is put into spectrum analysers where its frequency is measured.
}
\end{figure}

\subsection{Achieving the ultimate quantum limits with SPADE}

We have seen that direct imaging is optimal when the sources' PSFs are well separated. 
In this section, we derive a nearly optimal measurement for all values of the overlap parameter $\kappa$.
We obtain analytical expressions for the error exponents for this particular measurement.
Such a measurement is defined by a continuous family of POVM elements,
\begin{align}\label{eq:measurement1}
M_1(\omega) & = \ket{e_1}\bra{e_1} \otimes \ket{\omega}\bra{\omega} \, , \\ 
M_2(\omega) & = (\openone -\ket{e_1}\bra{e_1} )\otimes \ket{\omega}\bra{\omega} \, .
\label{eq:measurement2}
\end{align}
Physically, these measurement operators correspond to inputting the light into a spatial demultiplexer capable of separating the modes $\ket{e_1}$ and $\ket{e_2}$, followed
by spectral analysis. 
That is, the spectrum of the fundamental mode $\ket{e_1}$ is contrasted against the spectrum of the higher order modes.  
We refer to this measurement strategy as SPADE-assisted spectroscopy.
In particular, if the PSF is Gaussian, an optimal set of modes for demultiplexing is the Hermite-Gaussian basis (or Laguerre-Gaussian). 
Projection of the electric field on this basis allows us to maximally separate the light of the star from that of the planet. 
These devices are already known to be optimal for other problems in quantum imaging~\cite{PhysRevX.6.031033,PhysRevLett.117.190801,PhysRevLett.117.190802,tsang2019resolving,PhysRevLett.97.163903,wicker2007interferometric}, and can be realised with an interferometric setup~\cite{tsang2019resolving}, a hologram~\cite{paur2016achieving}, or a multimode wave guide~\cite{PhysRevX.6.031033}.
A schematic of the waveguide implementation is shown in Fig.~\ref{f:measurement}. 
Recall that a state of a single photon in the Hermite-Gaussian mode of order $q$ is 
\begin{align}
\ket{\phi_q} &= \int dx \, \phi_q(x) \, a^\dagger(x) \, , \qquad q = 0,1,... \nn
\phi_q(x) &= \frac{1}{2\pi \sigma^2}\frac{1}{\sqrt{2^q q!}} \, H_q\left(\frac{x}{\sqrt2\sigma}\right)
e^{-\frac{x^2}{4\sigma^2}} \, ,
\end{align}
and $H_q$ is the Hermite polynomial.

SPADE devices have recently been built and demonstrated~\cite{boucher2020spatial,fontaine2019laguerre}, with access to more than $200$ modes achieved with relatively simple experimental complexity~\cite{fontaine2019laguerre}. 
However, in contrast to performing source separation estimation \cite{PhysRevX.6.031033}, which in principle requires us to individually address a large number of higher order modes, here we only need to split the light coupled into the fundamental mode.
It is essential to align the optical imaging system with the centre of the star, therefore we have $|\phi_0\rangle = |\psi_0\rangle$, and $M_1(\omega) = \ket{\phi_0}\bra{\phi_0} \otimes \ket{\omega}\bra{\omega}$.
If the two sources are close enough, compared with the Rayleigh length $\sigma$, the Hermite-Gaussian modes of degree larger than $1$ are suppressed. This means that the operator is Eq.~(\ref{eq:measurement2}) is well approximated as $M_2(\omega) \simeq \ket{\phi_1}\bra{\phi_1} \otimes \ket{\omega}\bra{\omega}$, and it suffices, in this case, to sort only the first order mode.

For the Chernoff bound, we observe that the optimisation of Eq.~\eqref{eq:ccb} over $t$ occurs at $t=1/2$, yielding
\begin{align}
\xi_\text{CCB}^\text{SPADE} &= \frac{\epsilon(1-|\kappa|^2) \zeta}{W \sqrt{W-\eta\Delta }} + O(\epsilon^2) \, , 
\end{align}
where
\begin{align}
\zeta := -W^{\frac{3}{2}}+W \sqrt{W-\eta\Delta }-\Delta  \sqrt{W-\eta  W}+\Delta  \sqrt{W} \, .
\end{align}
Similarly, the relative entropy of the measurement in Eqs.~\eqref{eq:measurement1}-\eqref{eq:measurement2} is
\begin{align}
\vartheta_\text{CRE}^\text{SPADE}
= \frac{\epsilon (1-|\kappa|^2) \Omega}{W}  
   +\frac{ \epsilon^2 \eta^2 |\kappa|^4  (W-\Delta )\Delta}{2 (W - \eta \Delta )^2}+ 
    O\left(\epsilon ^3\right) \, , 
\label{eq:cre_opt}
\end{align}
with $\Omega$ as in Eq.~(\ref{Omegadef}).
To first order, this matches the ultimate quantum limit expressed by the quantum relative entropy in Eq.~(\ref{QRE-s}), 
and a small difference appears in the second order.

Figure~\ref{f:qre_vs_cre_main} shows the performance of SPADE-assisted spectroscopy for both symmetric and asymmetric hypothesis testing.
The figure compares SPADE with the ultimate quantum limits and with direct imaging. 
Indeed, this measurement strategy nearly saturates the ultimate quantum limits in both cases across all ranges of values of $|\kappa|^2$ (the difference is smaller than the size of the markers).

\subsection{Finite-size effects for the asymmetric setting}\label{sec:finite}

Up to now, we have computed the asymptotic error rates in the limit that $n\rightarrow \infty$.
We complete this work by considering finite-size corrections to the asymptotic error exponent, within the framework of asymmetric hypothesis testing.
Higher-order corrections of the quantum Stein lemma are necessary for finite-size analysis.
A revised version of the quantum Stein lemma holds 
for finite values of $n$, which includes corrections of order $1/\sqrt{n}$~\cite{tomamichel2013hierarchy,li2014second},
\begin{align}
\vartheta & \leq 
\vartheta_\text{QSL} \\ 
& := D(\rho_0 \| \rho_1) + \sqrt{\frac{V(\rho_0 \| \rho_1)}{n}} \, \Phi^{-1}(\alpha) + O\left( \frac{\log n}{n} \right) \, .
\label{eq:asymm}
\end{align}
In this latter expression, $\Phi^{-1}$ is the inverse of the error function $\Phi$, where
$\Phi(z) = 2/\sqrt{\pi}\int_0^z e^{-t^2} dt$, and $V(\rho_0||\rho_1)$ is the variance of the logarithmic likelihood ratio,
\begin{align}
V(\rho_0 \| \rho_1) = \text{Tr}[\rho_0 (\log\rho_0 - \log\rho_1)^2] - D(\rho_0 \| \rho_1)^2 \, .
\end{align}
Indeed, in realistic scenarios, the experimenter may be able to obtain only a limited number of photons, and therefore it is desirable to have bounds on the error probabilities in this regime.  

\begin{figure}[t!]
\includegraphics[trim = 0cm 0.0cm 0cm 0cm, clip, width=1.0\linewidth]{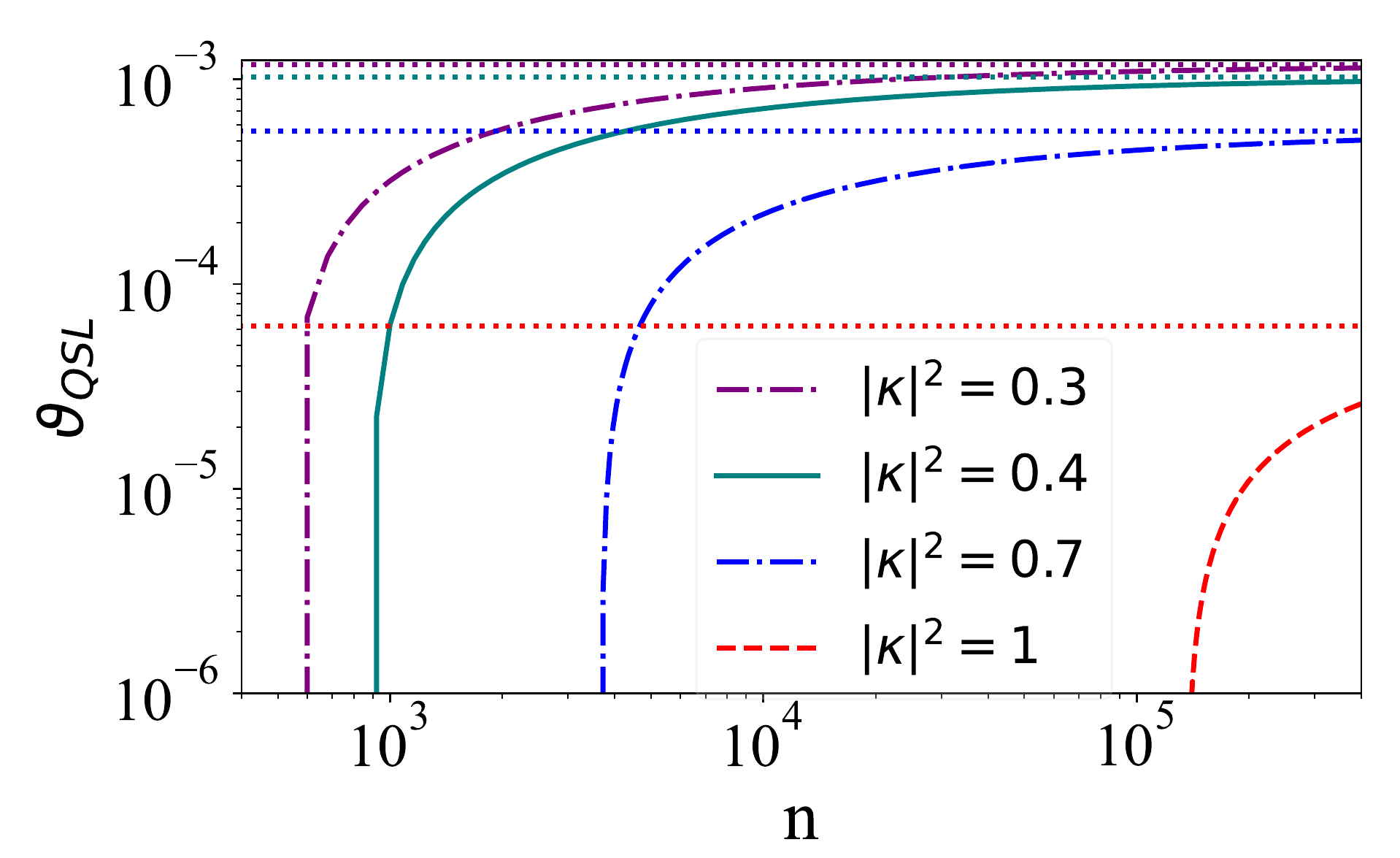} 
\caption{\label{f:asym_finite_n} The error exponent for asymmetric hypothesis testing, plotted vs the number of detected photons $n$, computed from Eq.~(\ref{eq:asymm}) for $\alpha = 0.05$, $\epsilon = 0.05$, $\eta = 0.5$, $W=1$, and $\Delta = 0.2 W$ shown in log-log scale.  We show $|\kappa|^2 = 0.3$ (purple dotted-dashed line), $|\kappa|^2 = 0.4$ (teal solid line), $|\kappa|^2 = 0.7$ (blue dotted-dashed line) and $|\kappa|^2 = 1$ (red dashed line). The straight horizontal lines indicate the asymptotic behaviour as $n\rightarrow \infty$, and the curved lines show $\vartheta_\text{QSL}$.
}
\end{figure}

We have seen that the leading term $D(\rho_0||\rho_1)$ increases linearly with $\epsilon$ and $(1-|\kappa|^2)$. 
However, the second order term, $\sqrt{V(\rho_0||\rho_1)} \Phi^{-1}(\alpha)$ is actually negative for all $\alpha \in (0,1/2]$, since $\Phi^{-1}(\alpha) < 0$.
In Fig.~\ref{f:asym_finite_n} we show the right-hand-side of Eq.~(\ref{eq:asymm}) as a function of $n$ (we neglect the correction term of order $O( n^{-1} \log n)$), computed for a Gaussian PSF.
Compared with the transit method, not only does the optimal quantum measurement achieve a higher error exponent,
but also with fewer photons.

\section{Discussion and conclusions}

From a broader perspective, our approach may be useful in any situation where the aim is to achieve selective spectroscopy of a pair of point-like sources with sub-Rayleigh separation.
Other authors have proposed methods for quantum-enhanced spectroscopy based on entanglement or squeezing \cite{QZhuang, casacio2021quantum, belsley2022quantum}.
These techniques require manipulating the objects or illuminating them with light that has special, non-classical properties. However, often it is the case, such as for astronomy, that we cannot illuminate the objects of interest. Rather all we can do is analyse the light that reaches us.

In this work, we have discussed how to implement optimal spectroscopy by applying a structured detection technique based on spatial demultiplexing, which does not require source engineering. A related approach has been discussed in the context of noise spectroscopy, but with demultiplexing in the frequency domain \cite{Ng2016,tsang2022quantum}.

Astronomical spectroscopy is a type of remote spectral sensing that is significant for our understanding of the universe.
We have established the ultimate quantum limit for determining the presence or absence of a spectral absorption line, for a dim source in the proximity of a much brighter stellar source.
We have characterised both the symmetric and asymmetric hypothesis testing error exponents and showed that the optimal quantum technique can significantly outperform both direct imaging and the transit method. Compared to direct imaging, we have shown that interferometric measurements in combination with spectrum analysis yield a quadratic improvement in the error exponent.
Intuitively, this {SPADE-assisted spectroscopy} outperforms direct imaging because it allows us to separate the light coming from the planet from that coming from the star.
The fundamental mode captures all of the parent star's light, which contains no useful information about the planet's atmosphere but overlaps spatially with the planet's mode and contributes a noisy background. 
By effectively rejecting the light coming from the star, SPADE can substantially improve the signal-to-noise ratio.
In the case of limited resources our theory allows the experimenter to determine quantitatively how many photons are required to achieve a certain error threshold, and hence determine whether a decision task is achievable in principle.

\begin{acknowledgments}
ZH is supported by a Sydney Quantum Academy Postdoctoral Fellowship.
\end{acknowledgments}



%


\appendix
\widetext

\counterwithin{figure}{section}
\section{The model}

\subsection{Transit spectroscopy}
\label{sec:transit}

Consider transit exoplanet spectroscopy, depicted in Fig.~\ref{f:schematic1} (a). The process follows
\begin{enumerate}
    \item We wait for the planet to transit in front of the star, signified by the fact that there is a decrease in the total intensity.
    \item The spectrum is examined and is compared to that of the star: we look for an absorption line in the spectrum.
\end{enumerate}

The density matrix is a linear combination of those of the star and planet, weighted by their intensities. 
Under $H_0$, the frequency spectrum is uniform
\begin{align}
\rho_0 = \frac{1}{W}\int_0^{W}d\omega \ket{\omega}\bra{\omega} \otimes \ket{\psi_0}\bra{\psi_0}.
\end{align}
Under $H_1$,
\begin{align}
\rho_1 &= (1-\epsilon) \rho_\text{star}+
                        \epsilon \rho_1^\text{planet} \nn
        &= \bigg[\left( \frac{1-\epsilon}{W} +\frac{\epsilon}{W-\eta\Delta} \right)\int_0^{\omega_0 - \Delta /2} d\omega \ket{\omega}\bra{\omega}+
          \left( \frac{1-\epsilon}{W} + \frac{\epsilon (1-\eta)}{W-\eta\Delta}    \right)\int_{\omega_0 - \Delta /2}^{\omega_0 + \Delta /2} d\omega \ket{\omega}\bra{\omega} +\nn
           &\left( \frac{1-\epsilon}{W} + \frac{\epsilon}{W-\eta\Delta} \int_{\omega_0 + \Delta /2}^W d\omega\ket{\omega}\bra{\omega}\right)\bigg] \otimes \ket{\psi_0}\bra{\psi_0}.
\end{align}

The two density matrices $\rho_0$ and $\rho_1$ are already diagonal in the basis of $\ket{\omega}\bra{\omega}$, and
we can compute the quantum Chernoff exponent and the quantum relative entropy straightforwardly.

For the QCE, we have

\begin{align}
\xi_\text{QCE}(\rho_0,\rho_1) = -\log  \bigg[ \min_{0\leq t \leq1} 
&(W-\eta\Delta )^{t-1} (\Delta  \eta  (\epsilon -1)+W)^{1-t}-\frac{\Delta  (W-\eta\Delta )^{t-1}
     (\eta\Delta(\epsilon -1)+W)^{1-t}}{W}+\nn
&\frac{\Delta  (W-\eta\Delta )^{t-1} (\eta\Delta (\epsilon -1)-\eta  W \epsilon +W)^{1-t}}{W}\bigg]
\end{align}

In the limit that $\epsilon \ll 1$, 

\begin{align}
\xi_\text{QCE} = \frac{\epsilon ^2 \eta ^2 \Delta   (W-\Delta )}{8 (W-\eta \Delta )^2}
                +O\left(\epsilon ^3\right).
\end{align}

In this case, the classical counterpart is the same as the quantum, since the density matrices are diagonal and correspond to measurement in the frequency basis.

\subsection{Spatially separate case}\label{AppA2}

Now, in the case where the star and the planet are spatially separated. Since now spatial degrees of freedom are involved, we need to assume a PSF for the imaging system, which we take to be real and Gaussian.
\begin{align}
\psi(x) = \left( \frac{1}{2\pi \sigma^2} \right)^{1/4} \exp\left(\frac{-x^2}{4\sigma^2}\right)
\end{align}

For this case, we have 2 degrees of freedom, spatial and frequency.

In the frequency degree of freedom, a photon from the star has spectral and spatial DOS, and  can be described as
\begin{align}
\rho_\text{star} = \frac{1}{W}\int^W d\omega \ket{\omega}\bra{\omega} \otimes \ket{\psi_{0}}\bra{\psi_{0}}.
\end{align}

The states of a photon coming from the planet under $H_0$ and $H_1$ are, respectively
\begin{align}
\rho^\text{planet}_0 &= \frac{1}{W}\int^W d\omega \ket{\omega}\bra{\omega} \otimes \ket{\psi_s}\bra{\psi_s}; \nn
\rho^\text{planet}_1  &= \frac{1}{W - \eta\Delta}
\left[\int_0^{\omega_0 - \Delta \omega/2} \ket{\omega}\bra{\omega} + (1-\eta) \int_{\omega_0-\Delta /2}^{\omega_0 + \Delta/2} \ket{\omega}\bra{\omega} + \int_{\omega_0 + \Delta/2} ^W  \ket{\omega}\bra{\omega} \right] \otimes \ket{
  \psi_{ s} } \bra{\psi_s } 
\end{align}

For $H_j, j \in \{0,1\}$, the density matrix is then
\begin{align}
\rho_j = (1-\epsilon) \rho_\text{star} + \epsilon \rho^j _\text{planet}.
\end{align}

To simplify the notation, we divide the frequency spectrum into three bands: we use $\omega_1 $ to denote the band 
$\omega \in [0,\omega_0- \Delta/2 ]$, $\omega_2 $ denoting $\omega \in [\omega_0- \Delta/2,\omega_0+ \Delta/2 ]$, and
$\omega_3$ for  $\omega \in [\omega_0+ \Delta/2, W ]$.

Let the star be positioned at $x_0$ and the planet at $x_0 +s$.

Under $H_0$, the outcomes are

\begin{align}
p_{H_0}(x,\omega_1) &=\frac{W-\Delta }{2 W} \left((1-\epsilon) |\psi(x-x_0)|^2 + \epsilon |\psi(x-x_0-s)|^2 \right) \nn
p_{H_0}(x,\omega_2) &=\frac{\Delta}{W} \left((1-\epsilon) |\psi(x-x_0)|^2 + \epsilon |\psi(x-x_0-s)|^2 \right) \nn
p_{H_0}(x,\omega_3) &=\frac{W-\Delta }{2 W} \left((1-\epsilon) |\psi(x-x_0)|^2 + \epsilon |\psi(x-x_0-s)|^2 \right)
\end{align}

Under $H_1$,
\begin{align}
p_{H_1}(x,\omega_1) &=\frac{\omega_0-\Delta/2}{W} (1-\epsilon) |\psi(x-x_0)|^2 + \frac{\omega_0-\Delta/2}{W-\eta \Delta}\epsilon |\psi(x-x_0-s)|^2  \nn
p_{H_1}(x,\omega_2) &=\frac{\Delta (1-\epsilon)}{W}  |\psi(x-x_0)|^2 + \epsilon \frac{\Delta(1-\eta)}{W-\eta \Delta}|\psi(x-x_0-s)|^2 \nn
p_{H_1}(x,\omega_3) &=\frac{W-\omega_0-\Delta/2}{W} (1-\epsilon) |\psi(x-x_0)|^2 + \epsilon  \frac{W-\omega_0-\Delta/2}{W-\eta \Delta}|\psi(x-x_0-s)|^2
\end{align}

The spatial distribution of $\ket{\psi_{x_0}}$ and $\ket{\psi_s}$ are, in general, not orthogonal. We can diagonalise these via 
\begin{align}
\ket{e_1} = \ket{\psi_{x_0}}, \qquad
\ket{e_2} = \frac{\ket{\psi_{x_0+s}} - \kappa\ket{\psi_{x_0}}}{\sqrt{1-\kappa^2}}
\end{align}

\noindent where we have defined $\kappa  = \braket{\psi_{x_0}|\psi_s}$. For a Gaussian PSF, $\kappa = \exp(-s^2/8\sigma^2);
$

To facilitate the calculation, we define density matrices in frequency space
\begin{align}
\rho^{\omega_1} &= \frac{1}{\omega_0-\Delta/2}\int_0^{\omega_0-\Delta/2} d\omega \ket{\omega}\bra{\omega}, 
\\
\rho^{\omega_2} &= \frac{1}{\Delta}\int_{\omega_0-\Delta/2}^{\omega_0+\Delta/2} d\omega \ket{\omega}\bra{\omega}, 
\\
\rho^{\omega_3} &= \frac{1}{W-(\omega_0+\Delta/2)}\int_{\omega_0+\Delta/2}^{W} d\omega \ket{\omega}\bra{\omega}.
\end{align}

The spectrum of the star can be written as
\begin{align}
\rho_\text{star}^\omega = \frac{1}{W}\left[ \bigg(\omega_0 - \frac{\Delta}{2}\bigg)\rho^{\omega_1} 
                                        + \Delta\rho^{\omega_2} +
                          \bigg(W-\omega_0-\frac{\Delta}{2}\bigg)\rho^{\omega_3}      \right]
\end{align}

The spectrum of the planet under $H_1$ can be written as
\begin{align}
\rho_\text{planet}^\omega = \frac{1}{W-\eta \Delta}\left[ \bigg(\omega_0 - \frac{\Delta}{2}\bigg)\rho^{\omega_1} + \Delta (1-\eta)\rho^{\omega_2} +
                          \bigg(W-\omega_0-\frac{\Delta}{2}\bigg)\rho^{\omega_3}               \right]
\end{align}

In the basis of $\{\ket{\omega} \} \otimes \{\ket{e_1},\ket{e_2} \}$, the density matrices are
\begin{align}  \label{eq:nondiag}
\rho_0 = & \rho_\text{star}^\omega \otimes ((1-\epsilon(1-\kappa^2))\ket{e_1}\bra{e_1}+ \epsilon (1-\kappa^2) \ket{e_2}\bra{e_2} +\epsilon \kappa\sqrt{1-\kappa^2} (\ket{e_1}\bra{e_2} +c.c) )\nn
\rho_1 =&\left((1-\epsilon) \rho^\omega_\text{star} + \epsilon \kappa^2\rho^\omega _\text{planet} \right)\otimes \ket{e_1}\bra{e_1} +
\epsilon (1-\kappa^2) \rho_\text{planet}^\omega \otimes \ket{e_2 }\bra{e_2} +\nn
&\epsilon \kappa \sqrt{1-\kappa} \rho_\text{planet}^\omega \otimes \ket{e_1}\bra{e_2} +c.c.
\end{align}


Under $H_0$, in the basis $\{\ket{e_1},\ket{e_2} \}\otimes \ket{\omega}$, the spatio-spectral domain components of the density matrices are

\begin{align}
\rho_{H_0}^{\omega_1} &= \left(
\begin{array}{cc}
 \frac{\kappa ^2 \epsilon  \left(\frac{W}{2}-\frac{\Delta }{2}\right)}{W}+\frac{(1-\epsilon ) \left(\frac{W}{2}-\frac{\Delta }{2}\right)}{W} & \frac{\kappa  \sqrt{1-\kappa ^2} \epsilon  \left(\frac{W}{2}-\frac{\Delta }{2}\right)}{W} \\
 \frac{\kappa  \sqrt{1-\kappa ^2} \epsilon  \left(\frac{W}{2}-\frac{\Delta }{2}\right)}{W} & \frac{\left(1-\kappa ^2\right) \epsilon  \left(\frac{W}{2}-\frac{\Delta }{2}\right)}{W} \\
\end{array}
\right) \otimes \int \ket{\omega}\bra{\omega}
\nn
\rho_{H_0}^{\omega_2} &= \left(
\begin{array}{cc}
 \frac{\Delta  \kappa ^2 \epsilon }{W}+\frac{\Delta  (1-\epsilon )}{W} & \frac{\Delta  \kappa  \sqrt{1-\kappa ^2} \epsilon }{W} \\
 \frac{\Delta  \kappa  \sqrt{1-\kappa ^2} \epsilon }{W} & \frac{\Delta  \left(1-\kappa ^2\right) \epsilon }{W} \\
\end{array}
\right) \otimes \int \ket{\omega}\bra{\omega} \nn
\rho_{H_0}^{\omega_3} &= \left(
\begin{array}{cc}
 \frac{\kappa ^2 \epsilon  \left(\frac{W}{2}-\frac{\Delta }{2}\right)}{W}+\frac{(1-\epsilon ) \left(\frac{W}{2}-\frac{\Delta }{2}\right)}{W} & \frac{\kappa  \sqrt{1-\kappa ^2} \epsilon  \left(\frac{W}{2}-\frac{\Delta }{2}\right)}{W} \\
 \frac{\kappa  \sqrt{1-\kappa ^2} \epsilon  \left(\frac{W}{2}-\frac{\Delta }{2}\right)}{W} & \frac{\left(1-\kappa ^2\right) \epsilon  \left(\frac{W}{2}-\frac{\Delta }{2}\right)}{W} \\
\end{array}
\right) \otimes \int \ket{\omega}\bra{\omega}
\end{align}

And we have
\begin{align}
\rho_0 = \rho_{H_0}^{\omega_1} \oplus \rho_{H_0}^{\omega_2} \oplus \rho_{H_0}^{\omega_3}
\end{align}

Now, under $H_1$, these are, instead
\begin{align}
\rho_{H_1}^{\omega_1}=
\begin{pmatrix}
(1-\epsilon)\frac{(\omega_0-\Delta/2)}{W} +  \epsilon \kappa^2\frac{(\omega_0 - \Delta/2)}{W- \eta \Delta} &
\epsilon \kappa \sqrt{1-\kappa^2} \frac{(\omega_0 - \Delta/2)}{W- \eta \Delta}\\
\epsilon \kappa \sqrt{1-\kappa^2} \frac{(\omega_0 - \Delta/2)}{W- \eta \Delta} &
\epsilon(1-\kappa^2)\frac{(\omega_0 - \Delta/2)}{W-\eta \Delta}
\end{pmatrix} \otimes \int_{0}^{\omega_0-\Delta/2} \ket{\omega}\bra{\omega}
\end{align}

\begin{align}
\rho_{H_1}^{\omega_2}= 
\begin{pmatrix}
(1-\epsilon)\frac{\Delta}{W} + \epsilon \kappa^2 \frac{\Delta (1-\eta)}{W-\eta \Delta} & 
\epsilon \kappa \sqrt{1-\kappa^2}\frac{\Delta (1-\eta)}{W-\eta \Delta} \\
\epsilon \kappa \sqrt{1-\kappa^2}\frac{\Delta (1-\eta)}{W-\eta \Delta}&
\epsilon (1-\kappa^2) \epsilon \kappa \sqrt{1-\kappa^2}\frac{W - \omega_0 -\Delta/2}{W-\eta \Delta}
\end{pmatrix} \int_{\omega_0-\Delta/2}^{\omega_0+\Delta/2} \ket{\omega}\bra{\omega}
\end{align}

\begin{align}
\rho_{H_1}^{\omega_3} =
\begin{pmatrix}
(1-\epsilon)\frac{W-\omega_0 - \Delta/2}{W} + \epsilon \kappa^2\frac{W-\omega_0 -\Delta/2}{W-\eta \Delta} &
\epsilon \kappa \sqrt{1-\kappa^2}\frac{W - \omega_0 -\Delta/2}{W - \eta \Delta} \\
\epsilon \kappa \sqrt{1-\kappa^2}\frac{W - \omega_0 -\Delta/2}{W - \eta \Delta}&
\epsilon (1-\kappa^2) \frac{W - \omega_0 -\Delta/2}{W - \eta \Delta}
\end{pmatrix} \int_{\omega_0+\Delta/2}^{W} \ket{\omega}\bra{\omega}
\end{align}

We have, under $H_1$
\begin{align}
\rho_1 = \rho_{H_1}^{\omega_1} \oplus \rho_{H_1}^{\omega_2} \oplus \rho_{H_0}^{\omega_3}
\end{align}

The quantum Chernoff exponent (QCE) is difficult to compute analytically due to the optimisation, but one can compute the QRE. 

The full expression for the QRE is cumbersome and not particularly enlightening. But in the limit $\epsilon \ll1$, the expression is

\begin{align}
D(\rho_0||\rho_1) &\approx \frac{\epsilon(1-\kappa^2)(W \log \left(1-\frac{\eta \Delta }{W}\right)
-\Delta  \log (1-\eta ))}{W} + \frac{\epsilon ^2 \kappa ^2 }{2 (W-\eta \Delta  )^2} \times 
\bigg[2 \bigg(-\eta ^2 \Delta ^2  \left(\kappa ^2-1\right) \log \left(\epsilon -\kappa ^2 \epsilon \right) \nn
&+\kappa ^2 W^2 \log \left(1-\frac{\eta \Delta }{W}\right)+2 \Delta ^2 \eta ^2 \kappa ^2 \log \left(\frac{W-\eta  \Delta }{W-\eta  W}\right)+2 \Delta ^2 \eta ^2 \log \left(\frac{W-\eta  W}{W- \eta \Delta}\right)-  \eta ^2 \Delta \kappa ^2 W \log (W- \eta \Delta)\nn
&+ \eta ^2 \Delta  W \log (W-\eta \Delta )+W \left(2 \eta  \Delta  \kappa ^2+W\right) \log \left(\frac{W}{W- \eta \Delta }\right)+2 \eta \Delta  W \log \left(1-\frac{\eta  \Delta }{W}\right)\nn
&+  \eta ^2  \Delta \kappa ^2 W \log \left((\eta -1) \left(\kappa ^2-1\right) W \epsilon \right)-\Delta  \eta ^2 W \log \left((\eta -1) \left(\kappa ^2-1\right) W \epsilon \right)\bigg)\nn
&+ \eta ^2 \Delta  \left(3 \kappa ^2-2\right) (W-\Delta )-2 \Delta  \left(\kappa ^2-1\right) \log (1-\eta ) (W-2   \eta  \Delta)
 \bigg] + O(\epsilon^3)
\end{align}
Note that the first-order in $\epsilon$ is the same as the first-order term for the almost-optimal measurement.

Now, let us examine the almost-optimal quantum measurement. The POVMs are
\begin{align}\label{eq:measurement_append}
\hat M(e,\omega) = \ket{e_1}\bra{e_1} \otimes \ket{\omega}\bra{\omega} + (\openone -\ket{e_1}\bra{e_1} )\otimes \ket{\omega}\bra{\omega}
\end{align}

Divide into the three frequency bands, and associate the light that ends up in the fundamental mode with the label $e_1$,
 and  $\bar e_1$ otherwise.

For $H_0$
\begin{align}
p_{H_0}(e_1,\omega_1) &=  \frac{(2 \omega_0-\Delta) \left(1-\left(1-\kappa ^2\right) \epsilon \right)}{2 W} \nn
p_{H_0}(\bar e_1,\omega_1) &= \frac{\left(\kappa ^2-1\right) \epsilon  (\Delta -2 \omega_0)}{2 W}  \nn
p_{H_0}(e_1,\omega_2) &= \frac{\Delta - \Delta  \left(1-\kappa ^2\right) \epsilon  }{W}\nn
p_{H_0}(\bar e_1,\omega_2) &= \frac{\Delta  \left(1-\kappa ^2\right) \epsilon }{W}\nn
p_{H_0}(e_1,\omega_3) &=  \frac{\left(1-\left(1-\kappa ^2\right) \epsilon \right) (-\Delta +2 W-2 \omega_0)}{2 W}\nn
p_{H_0}(\bar e_1,\omega_3) &= \frac{\left(1-\kappa ^2\right) \epsilon  \left(-\frac{\Delta }{2}+W-\omega_0\right)}{W}
\end{align}

For $H_1$
\begin{align}
p_{H_1}(e_1,\omega_1) &= \left(\omega_0-\frac{\Delta }{2}\right) \left(\frac{\kappa ^2 \epsilon }{W-\eta \Delta }+\frac{1-\epsilon }{W}\right) \nn
p_{H_1}(\bar e_1,\omega_1) &= \frac{\left(\kappa ^2-1\right) \epsilon  (\Delta -2 \omega_0)}{2 (W-\eta \Delta)} \nn
p_{H_1}(e_1,\omega_2) &= \frac{\Delta  (1-\eta ) \kappa ^2 \epsilon }{W-\eta \Delta }+\frac{\Delta -\Delta  \epsilon }{W}\nn
p_{H_1}(\bar e_1,\omega_2) &= \frac{\Delta  (1-\eta ) \left(1-\kappa ^2\right) \epsilon }{W-\eta \Delta } \nn
p_{H_1}(e_1,\omega_3) &= \frac{(-\Delta +2 W-2 \omega_0) \left(\eta \Delta(\epsilon -1)+\left(\kappa ^2-1\right) W \epsilon +W\right)}{2 W (W-\eta \Delta)} \nn
p_{H_1}(\bar e_1,\omega_3) &= \frac{\left(1-\kappa ^2\right) \epsilon  \left(-\frac{\Delta }{2}+W-\omega_0\right)}{W-\Delta  \eta }
\end{align}

\subsubsection{Quantum Chernoff boud}
For the symmetric case, the QEC is difficult to compute analytically, but it can be calculated numerically. The classical Chernoff exponent VS the QCE is plotted in the main body of the manuscript. Qualitatively the error exponent behaves very similarly to the asymmetric case (see below).

The CCE of SPADE is given by

\begin{align}
\xi_\text{CCB}^\text{SPADE} = -\log\Biggl[& -\Delta  \left(\kappa ^2-1\right) \epsilon  \sqrt{\frac{1-\eta }{W^2-\eta \Delta   W}} 
-\frac{\left(\kappa ^2-1\right) \epsilon  (W-\Delta )}{\sqrt{W (W-\eta \Delta  )}} \nn
&+\sqrt{\frac{\left(\Delta  \left(\kappa ^2-1\right) \epsilon +\Delta \right) \left(\frac{\Delta  (\eta -1) \kappa ^2 \epsilon }{\eta \Delta  -W}+\frac{\   Delta -\Delta  \epsilon }{W}\right)}{W}}
 +\frac{(W-\Delta ) \sqrt{\frac{\left(\left(\kappa ^2-1\right) \epsilon +1\right) \left(\eta \Delta   (\epsilon -1)+\left(\kappa ^2-1\right) W \epsilon +W\right)}{W- \eta \Delta  }}}{W}
\Biggr] \\
&\approx  \frac{\left(1-\kappa ^2\right) \epsilon  \left(-W^{3/2}+W \sqrt{W-\eta \Delta }-\Delta  \sqrt{W-\eta  W}+\Delta  \sqrt{W}\right)}{W \sqrt{W-\eta \Delta}} + O(\epsilon^2)
\end{align}
in the limit that $\epsilon \ll1 $.


\subsubsection{Quantum Stein lemma}

For the CRE of the almost-optimal measurement, the expression is

\begin{align}
D_c(p_{H_0}^\text{SPADE}||p_{H_1}^\text{SPADE}) =&\log \left(\left(\left(\kappa ^2-1\right) \epsilon +1\right) (W-\eta \Delta )\right)+\frac{\Delta  \kappa ^2 \epsilon  \log ((1-\eta ) X)}{W}-\frac{\Delta  \epsilon  \log ((1-\eta ) X)}{W}+\frac{\Delta  \log (X)}{W}\nn
&-\frac{\Delta  \kappa ^2 \epsilon  \log (Y)}{W}+\frac{\Delta  \epsilon  \log (Y)}{W}-\frac{\Delta  \log (Y)}{W}+\kappa ^2 \epsilon  \log \left(\left(\kappa ^2-1\right) W \epsilon +W\right) \nn
&-\epsilon  \log \left(\left(\kappa ^2-1\right) W \epsilon +W\right)+\kappa ^2 (-\epsilon ) \log (X)+\epsilon  \log (X)-\log (X)\nn
X &\equiv \eta \Delta  (\epsilon -1)+\left(\kappa ^2-1\right) W \epsilon +W \nn
Y &\equiv \eta \Delta  (\epsilon -1)-W \left((\eta -1) \kappa ^2 \epsilon +\epsilon -1\right)
\end{align}

In the limit that $\epsilon \ll1$,
\begin{align}
D = \frac{\epsilon\left(\kappa ^2-1\right)   (\Delta  \log (1-\eta )-W \log (W-\eta \Delta  )+W \log (W))}{W} +
\frac{ \epsilon ^2 \eta ^2 \kappa ^4  (W-\Delta )\Delta}{2 (W- \eta \Delta )^2} +
 O\left(\epsilon ^3\right)
\end{align}


\section{Asymmetric error hypothesis testing}
\label{sec:qre_var}

Suppose the type-I error is no larger than a constant $\alpha$, then the type-II error exponent can be written as
\begin{align}
\beta =N D(\rho_0||\rho_1) + \sqrt{N V(\rho_0||\rho_1)} \Phi^{-1}(\alpha) + O(\log N)
\end{align}
where $\Phi^{-1}(y) = 1/\sqrt{2\pi}\int_{-\infty}^y ~dx \exp(-x^2/2).$ For $\alpha \in (0,1/2],~ \Phi^{-1}(\alpha) <0$, and $D(\rho_0||\rho_1)$ is the quantum relative entropy.
\begin{align}
D(\rho_0||\rho_1) &= \tr[\rho_0 (\ln(\rho_0 -  \rho_1))] \nn
                  &= \sum_i p_i \log p_i - \sum_i \braket{i| \rho_0 \log \rho_1 |i} \nn
                  &= \sum_i p_i (\log p_i - \sum_j P_{ij} \log q_j) \nn
\braket{i|  \log \rho_1 |i} &= \sum_j \log(q_j) P_{ij}, \qquad P_{ij} = \braket{i|j}\braket{j|i}
\end{align}

\noindent
Calculate the QRE variance. We have $\rho = \sum_i p_i\ket{i}\bra{i}$ and $\log\rho = \sum_i \log p_i \ket{i}\bra{i}$.

\begin{align}
V(\rho||\sigma) &= \text{Tr}[\rho (\log\rho - \log\sigma)^2] - D(\rho||\sigma)^2 
\end{align}
\begin{align}
\text{Tr}[\rho (\log\rho - \log\sigma)^2] &= \text{Tr}[\rho (\log\rho - \log\sigma)(\log\rho - \log\sigma)] \nn
&=\text{Tr}[\rho \left([\log\rho]^2 -\log\rho\log\sigma - \log\sigma\log\rho +[\log\sigma]^2  \right)]      \nn
\end{align}

\noindent
The cross-terms are

\begin{align}
\log\rho \log\sigma &= \sum_{i,j} \log p_i \log q_j \ket{i}\braket{i|j}\bra{j}\nn
\rho \log\rho \log\sigma &= \left(\sum_k p_k \ket{k}\bra{k}\right)\left(\sum_{i,j} \log p_i \log q_j \ket{i}\braket{i|j}\bra{j}\right) \nn
&= \sum_{i,j} p_i \log p_i \log q_j \ket{i}\braket{i|j}\bra{j}
\end{align}
Taking the trace,
\begin{align}
\bra{i}\sum_{i,j} p_i \log p_i \log q_j \ket{i}\braket{i|j}\braket{j|i} 
&=p_i \log p_i \log q_j P_{i,j}
\end{align}

\noindent
For the last term, we have
\begin{align}
\text{Tr}[\rho \log^2 \sigma] &= \text{Tr}\left[\sum_i p_i\ket{i}\bra{i}\left(\sum_j\log^2 q_j \ket{j}\bra{j}   \right) \right] \nn
&=\sum_{i,j}p_i \log^2 q_j P_{i,j}
\end{align}

\noindent
Therefore
\begin{align}
V(\rho||\sigma) &= \text{Tr}[\rho (\log\rho - \log\sigma)^2]
\end{align}
\begin{align}
\text{Tr}[\rho (\log\rho - \log\sigma)^2] 
&=\sum_i p_i \log^2 p_i - 2\sum_{i,j}p_i \log p_i \log q_j P_{i,j} + \sum_{i,j}p_i \log^2 q_j P_{i,j}
\end{align}

\end{document}